\documentclass[iop]{emulateapj}
\usepackage[utf8]{inputenc}
\usepackage[english]{babel}
\usepackage{natbib}
\usepackage{amsmath}
\usepackage{comment}
\usepackage{gensymb}
\usepackage{color}
\usepackage[hidelinks]{hyperref}
\begin{document}

\newcommand{\change}[1]{#1}

\newcommand{\tauboo}{\ensuremath{\tau} Boo}
\newcommand{\hatpeleven}{HAT-P-11}
\newcommand{\hatpelevenb}{HAT-P-11b}
\newcommand{\keplerthirteen}{Kepler-13}
\newcommand{\keplerthirteenb}{Kepler-13b}
\newcommand{\keplerseventeen}{Kepler-17}
\newcommand{\keplerseventeenb}{Kepler-17b}
\newcommand{\macula}{\texttt{macula}}
\newcommand{\kepler}{\textit{Kepler}}
\newcommand{\strobo}{\textit{stroboscopic effect}}
\newcommand{\rmeffect}{Rossiter--Mc\-Laugh\-lin effect}
\newcommand{\apriori}{\textit{a priori}}
\newcommand{\totalspots}{203}

\title{Stellar rotation--planetary orbit period commensurability in the \hatpeleven{} system}
\author{Bence Béky\altaffilmark{1},
Matthew J.~Holman\altaffilmark{1},
David M.~Kipping\altaffilmark{1,2},
Robert W.~Noyes\altaffilmark{1}}
\shortauthors{Béky et al.}
\altaffiltext{1}{Harvard--Smithsonian Center for Astrophysics, 60 Garden St, Cambridge, MA 02138, USA}
\altaffiltext{2}{Carl Sagan Fellow}
\email{bbeky@cfa.harvard.edu}

\begin{abstract}
A number of planet-host stars have been observed to rotate with a period equal to an integer multiple of the orbital period of their close planet. We expand this list by analyzing \kepler{} data of \hatpeleven{} and finding a period ratio of 6:1. In particular, we present evidence for a long-lived spot on the stellar surface that is eclipsed by the planet in the same position four times, every sixth transit. We also identify minima in the out-of-transit lightcurve and confirm that their phase with respect to the stellar rotation is mostly stationary for the 48-month timeframe of the observations, confirming the proposed rotation period. For comparison, we apply our methods to \keplerseventeen{} and confirm the findings of \citet{2012A&A...547A..37B} that the period ratio is not exactly 8:1 in that system. Finally, we provide a hypothesis on how interactions between a star and its planet could possibly result in an observed commensurability for systems where the stellar differential rotation profile happens to include a period at some latitude which is commensurable to the planetary orbit.
\end{abstract}


\section{Introduction}
\label{sec:introduction}

\change{Many stars} have been observed to exhibit photometric variations synchronous to the orbit of their close planet. When these variations are attributed to photospheric features rotating with the stellar surface, this implies a synchronicity between stellar rotation and planetary orbit. One of the earliest robust detections of this phenomenon is by \citet{2008A&A...482..691W} in the system \tauboo{}. They report on periodic photometric variations of the host star with a period within 0.04\% of that of the planetary orbit, and attribute this to an active region on the surface of the star. Similarly, stellar photometric variations synchronous to the planetary orbit have been detected for the planetary systems CoRoT-2 \citep{2009EM&P..105..373P,2009A&A...493..193L} and CoRoT-4 \citep{2009A&A...506..255L}. For all three stars, the rotation period inferred from spectroscopy is consistent with the period of photometric variations, supporting that the variations are due to photospheric features stationary on the stellar surface.

Another interesting example is \keplerthirteen{}. \citet{2012MNRAS.421L.122S} measure the rotational period of the star by frequency analysis of the spot-modulated lightcurve and find a 5:3 commensurability with the orbital period of the planet \keplerthirteenb{} at high significance. 

However, frequency analysis is not the only method suitable for measuring rotation rates of spots on the stellar surface. A transiting planet may eclipse spots on the surface of its host star, resulting in anomalies in the transit lightcurve. This phenomenon was observed, for example, in the systems \change{HD 209458 \citep{2003ApJ...585L.147S},} HD 189733 \citep{2007A&A...476.1347P}, TReS-1 \citep{2009A&A...494..391R}, and CoRoT-2 \citep{2009A&A...493..193L}. Repeated transit anomaly detections due to the same spot can be used to constrain the stellar rotation period. This method was first applied by \citet{2008ApJ...683L.179S} to HD 209458.

\change{Another application of starspot-induced transit anomalies is to constrain the spin-orbit geometry, as was first mentioned by \citet{2010ApJ...723L.223W}.}  This method was developed and applied independently by \citet{2011ApJ...740...33D} and \citet{2011ApJ...743...61S} to \hatpeleven{}, by \citet{2011ApJ...733..127S} to WASP-4, and by \citet{2011ApJ...740L..10N} to CoRoT-2.

Independent measurements of the \rmeffect{} on \hatpeleven{} show that the planetary orbit normal is almost perpendicular to the projected stellar spin \citep[the projected obliquity is $\approx103\degree$, see][]{2010ApJ...723L.223W,2011PASJ...63S.531H}. Relying only on photometric data, \citet{2011ApJ...740...33D} and \citet{2011ApJ...743...61S} independently identify two active latitudes (where spots are most prevalent) on the surface of the star, which they assume to be symmetrical around the equator, to conclude that \hatpelevenb{} is on a nearly polar orbit in accordance with the spectroscopic results, and that the stellar spin axis of \hatpeleven{} is close to being in the plane of sky.

\change{The transit lightcurve of \keplerseventeenb{} ($P=1.49\;\mathrm{days}$) also exhibits anomalies due to spots on the surface of its host star.  In their discovery paper, \citet{2011ApJS..197...14D} analyze these anomalies to study both stellar rotation and orbital geometry. They observe that the transit anomaly pattern repeats every eighth planetary orbit, suggesting that the spots rotate once while the planet orbits eight times. They dub the phenomenon of the same spots reappearing periodically at the same phase in transit lightcurves---every eighth one in this case---the \strobo{}.  As for the orbital geometry, they found that transit anomalies in successive orbits are consistent with being caused by the same spots that rotate one eighth of a full revolution on the stellar surface with each orbit of the planet.  This implies a low projected obliquity of the planetary orbit, and also excludes frequency aliases (like the star rotating three or five times while the planet orbits eight times).}

In this paper, we present evidence for a 6:1 period commensurability between the rotation of the star \hatpeleven{} and the orbit of its planet \hatpelevenb{} \citep[$P=4.89\;\mathrm{days}$,][]{2010ApJ...710.1724B}. The increasing number of systems known to exhibit such commensurability raises the question whether this is the result of an interaction between the planet and the star.

Whenever studying stellar rotation, it is important to remember that stars with convective zones exhibit differential rotation. In this paper, the working definition of stellar rotation rate is that inferred through dominant spots on the stellar surface, either from the rotational modulation of the out-of-transit lightcurve or from transit anomalies. This way we measure the rotation rate of the stellar surface at the latitude of the spots or active regions. If spots from multiple latitudes with different rotational rates contribute significantly to the lightcurve, then we expect the inferred posterior distribution of the rotational period to have a broader profile.

Despite their usefulness in confining planetary obliquity and mapping spots, transit anomalies due to the planet eclipsing spots can be a nuisance too: they contaminate the transit lightcurve, introducing biases in the detected transit depth \citep{2009A&A...505.1277C}, time, and duration. \citet{2010A&A...520A..66B} point out that in the particular case of stellar rotation-planetary orbit commensurability, activity-induced transit timing variations (TTVs) can be periodic, and thus can result in spurious planet detections. This further motivates the need for understanding stellar rotation--planetary orbit commensurability.

In Section \ref{sec:acf}, we look at the periodogram and autocorrelation function of \hatpeleven{} and \keplerseventeen{} lightcurves to confine the rotational period. In Section \ref{sec:anomalies}, we present the case of a spot on \hatpeleven{} recurring multiple times due to the \strobo{}. In Section \ref{sec:macula}, we analyze all transit anomalies observed on \hatpeleven{} to feed the best-fit spot parameters into the rotational modulation model \macula{} \citep{2012MNRAS.427.2487K}, and compare the resulting model lightcurve to observations. In Section \ref{sec:recurrence}, we perform a statistical analysis of spot-induced anomalies in the transits of \hatpelevenb{} and \keplerseventeenb{}. In Section \ref{sec:flipflop}, we look for the periodicity of lightcurve minima for both stars. We show evidence for two spots or spot groups at opposite longitudes on both \hatpeleven{} and \keplerseventeen{}, and find that on the former, they seem to alternate in relative activity level, which is known as the ``flip-flop'' phenomenon \change{\citep{1991LNP...380..381J}}. In Section \ref{sec:interaction}, we state one possible hypothesis about stellar rotation--planetary orbit resonance, and discuss difficulties in proving it. Finally, we summarize our findings in Section \ref{sec:conclusion}.

\section{Out-of-transit lightcurve}
\label{sec:acf}

In their discovery paper, \citet{2010ApJ...710.1724B} report a strong frequency component in the HATNet lightcurve of \hatpeleven{} with a period of approximately 29.2 days. They attribute it to rotational modulation of starspots, noting that the 6.4 mmag amplitude is consistent with observations of other K dwarfs, and the period is consistent with the color, activity level, and projected rotational velocity of \hatpeleven. They also note that both the secondary peaks in the autocorrelation function and the phase coherence of the lightcurve indicate that starspots (or spot groups) persist ``for at least several rotations''.

\begin{figure}
\begin{center}
\includegraphics*[width=89mm]{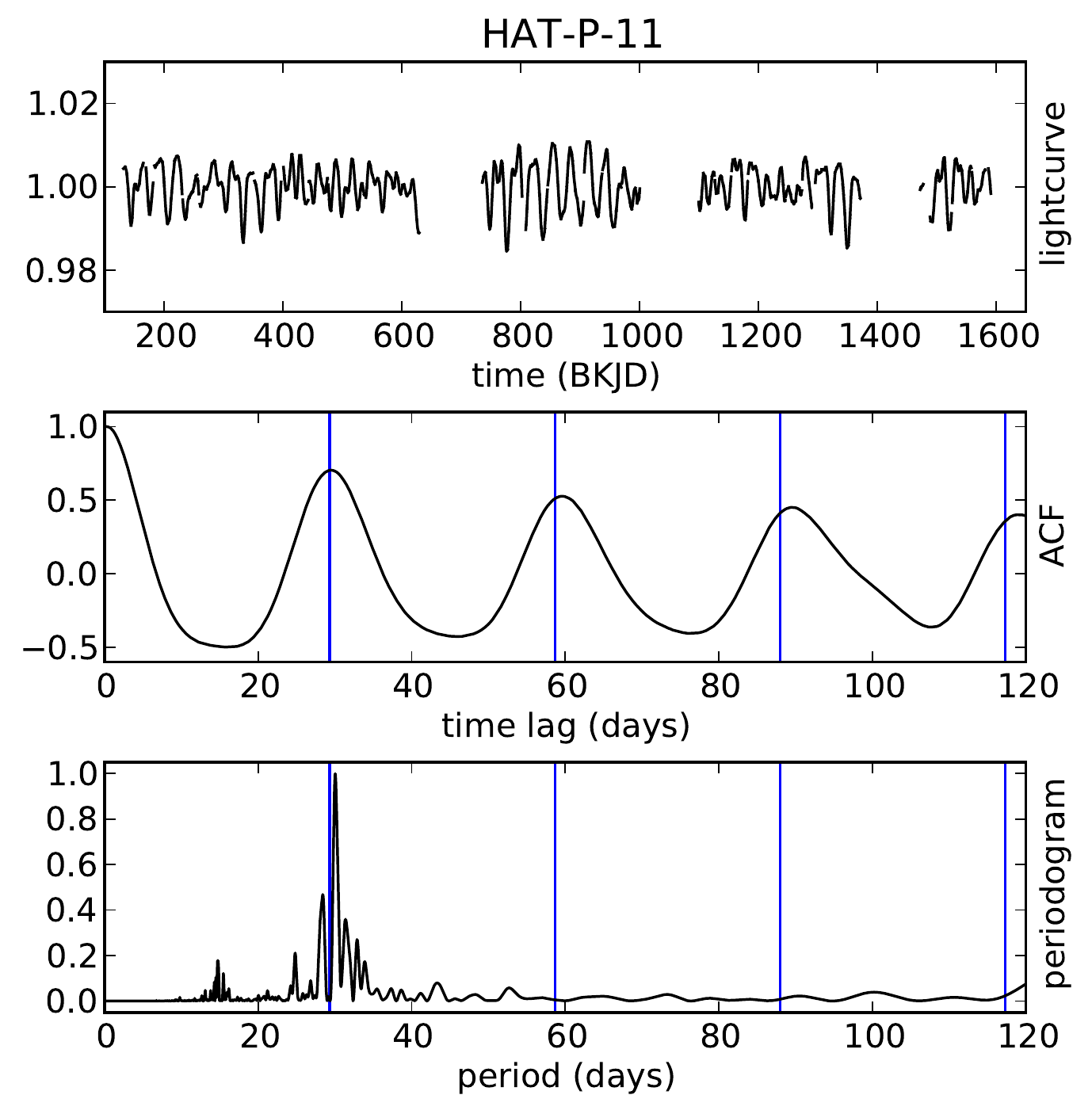}
\end{center}
\caption{Top panel: long cadence \kepler{} lightcurve of \hatpeleven, with transits of \hatpelevenb{} removed. Middle panel: autocorrelation function of the lightcurve. Bottom panel: Lomb--Scargle periodogram of the lightcurve. The blue vertical lines on the middle and bottom panels correspond to the proposed rotational period (six times the planetary orbital period), and its integer multiples.}
\label{fig:acf0}
\end{figure}

Figure \ref{fig:acf0} presents the entire \kepler{} space telescope \citep{2010Sci...327..977B} long cadence lightcurve of \hatpeleven{} (quarters 0--6, 8--10, 12--14, and 16--17, with transits of \hatpelevenb{} removed, and each quarter scaled to have unit mean). Time is measured in Barycentric \kepler{} Julian Date (BKJD), which is $\textrm{BJD}_\mathrm{UTC}-2\,454\,833$. Figure \ref{fig:acf0} also displays the autocorrelation function and periodogram of the long cadence lightcurve. This analysis \change{is similar to that} performed by \citet{2010ApJ...710.1724B}, but on much better quality data. We confirm their findings: we identify a peak in the autocorrelation function at a timelag of 29.32 days (with FWHM 8.05 days), and a peak in the periodogram at 30.03 days (with FWHM 0.62 days), \change{which we identify with the rotational period of \hatpeleven{}}. For comparison, six times the planetary orbital period is 29.33 days, and it is indicated along with its integer multiples on Figure \ref{fig:acf0} by blue vertical lines. We also see multiple peaks in the autocorrelation function at integer multiples of the base period, indicating that some spots must live for multiple stellar rotations.

\begin{figure}
\begin{center}
\includegraphics*[width=89mm]{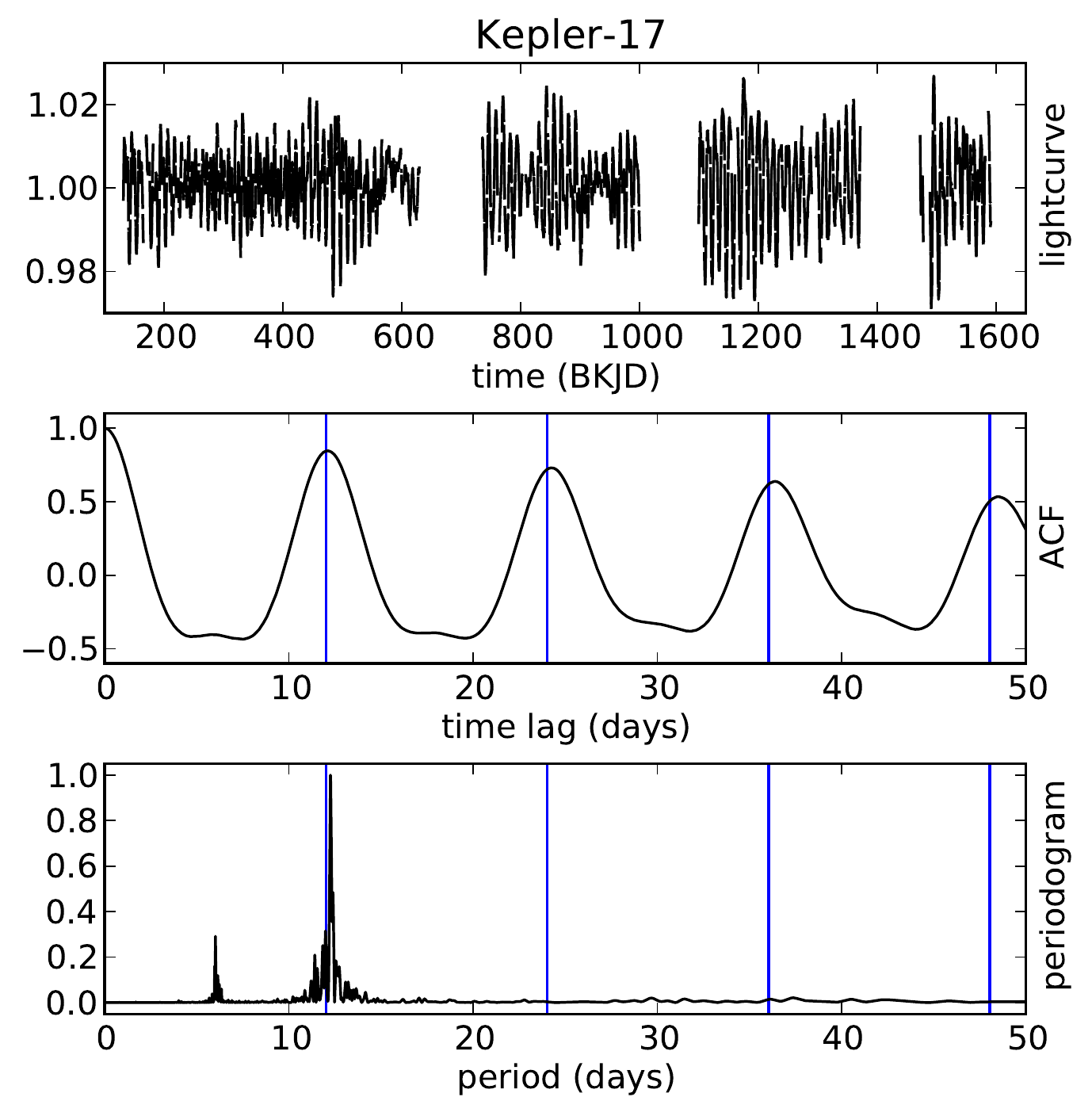}
\end{center}
\caption{Same as Figure \ref{fig:acf0}, for \keplerseventeen, with transits of \keplerseventeenb{} removed. The blue vertical lines on the middle and bottom panel correspond to the rotational period proposed by \citet{2012A&A...547A..37B} (not exactly eight times the planetary orbital period), and its integer multiples.}
\label{fig:acf1}
\end{figure}

For comparison, on Figure \ref{fig:acf1} we present the same analysis for the \kepler{} long cadence lightcurve of \keplerseventeen{} (quarters 1--6, 8--10, 12--14, and 16--17, with transits of \keplerseventeenb{} removed, and each quarter scaled to have unit mean).  The blue vertical lines indicate multiples of 12.01 days, the stellar rotation period reported by \citet{2012A&A...547A..37B}, instead of eight times the planetary orbital period, which is 11.89 days. The first peak of the autocorrelation function is at 12.10 days (with FWHM 3.13 days), while the periodogram peaks at 12.25 days (with FWHM 0.11 days). It is interesting to note that \hatpeleven{} and \keplerseventeen{} are in the same \kepler{} subfield on the sky, therefore we see gaps in both lightcurves during quarters 7, 11, and 15 due to the failure of a readout module.

The main reason for studying the autocorrelation function and the periodogram is to exclude the possibility of frequency aliases. If we interpret the half width of half maximum of the autocorrelation function as a direct indicator for period uncertainty \citep[as done, for example, by][]{2008A&A...488L..43A}, the resulting range is consistent with the proposed rotational periods for both stars. We refer the reader to \citet{2013MNRAS.432.1203M} for a discussion of using the autocorrelation function as a complementary method to periodograms for studying stellar rotation.

\change{Note that Kepler data are dense in time, with long runs of almost continuous observations.  We confirm that the spectral window function does not have large values at periods above 30 minutes, the cadence of observations.  Therefore unlike for sparsely sampled ground based observations, frequency aliasing \citep{2010ApJ...722..937D} does not pose a problem in this analysis.}

The periodograms \change{rule out} that we are dealing with an alias of the rotational rate. However, the narrow periodogram peak is \change{located at a period} slightly larger than the proposed rotational period for both stars. \citet{2013MNRAS.432.1203M} observe that spot evolution and differential rotation can cause periodogram peaks to split up into multiple narrow peaks, thus the FWHM may not correspond directly to the period uncertainty. Therefore the periodograms are not inconsistent with the proposed rotational periods of 29.33 days for \hatpeleven{} and 12.01 days for \keplerseventeen{}.

\section{Stroboscopic effect on \hatpeleven{}}
\label{sec:anomalies}

\citet{2010ApJ...723L.223W} were the first to note that the ratio between the stellar rotation period of \hatpeleven{} and the orbital period of \hatpelevenb{} is approximately 6:1. If it was close enough to 6:1, and there were spots that lived long enough, then one would be able to detect multiple lightcurve anomalies due to the same spot every sixth transit. However, \hatpelevenb{} has a polar orbit with respect to the stellar spin axis, therefore if the periods were incommensurable, then the spot could not fall repeatedly under the transit chord.

\citet{2011ApJ...743...61S} pointed out that a 6:1 period ratio is \apriori{} unlikely. They were looking for recurrence of transit anomalies, but quarters 0, 1, and 2 of \kepler{} data available at the time did not provide a large enough sample for such investigations.

\begin{figure}
\begin{center}
\includegraphics*[width=89mm]{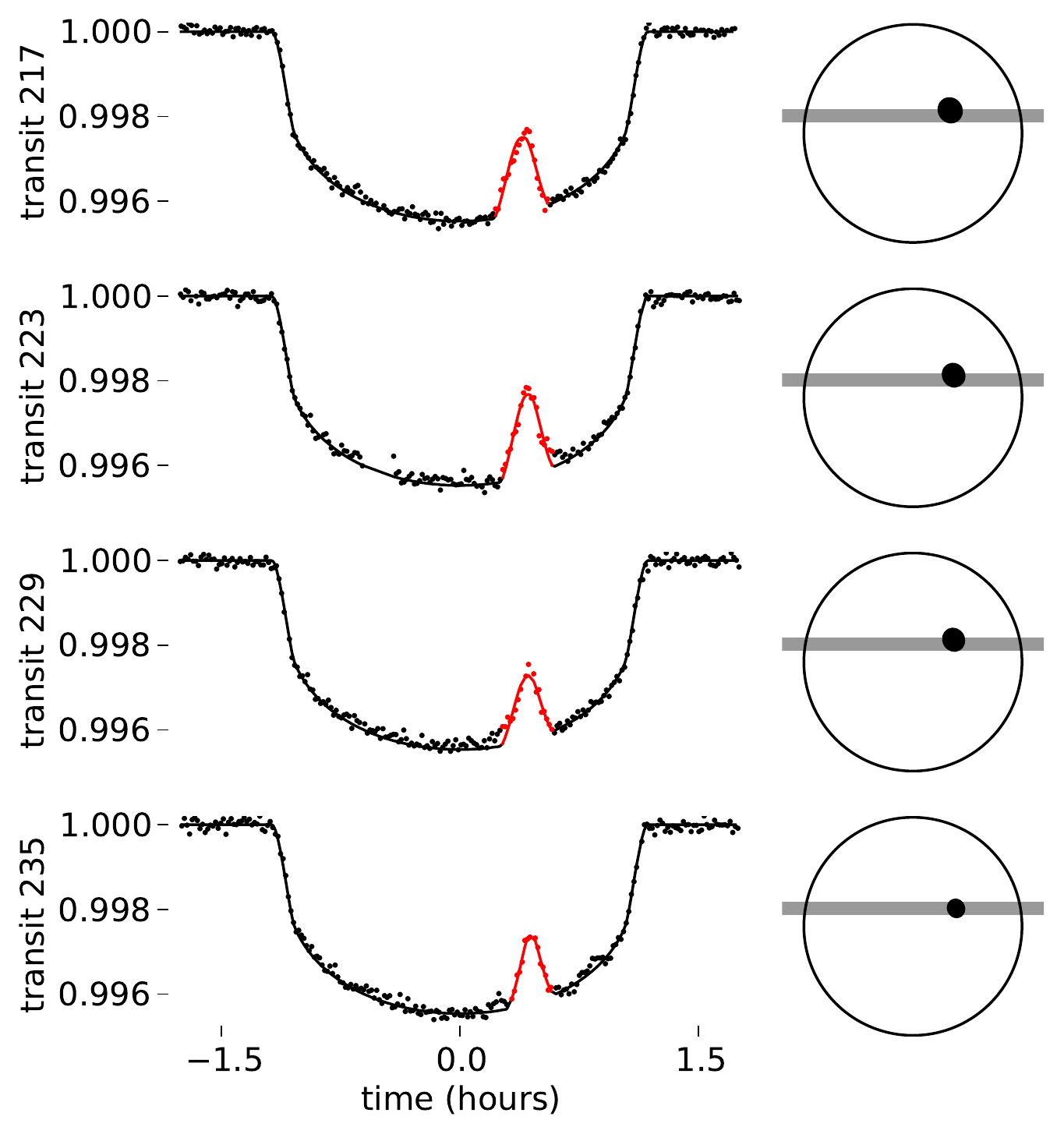}
\end {center}
\caption{Transit anomalies providing evidence for the 6:1 commensurability. The transits, from top to bottom, are separated by six planetary orbits, which is the proposed stellar rotation period. Left panels show detrended \kepler{} short cadence photometry, along with best fit model lightcurve with single spot. Right panels show the projected stellar disk, transit chord, and best fit spot. Note that spot seems to be stationary over this time period, which suggests a tight 6:1 commensurability.}
\label{fig:recurrence}
\end{figure}

In this section, we study a single extraordinary example of spot recurrence observed by the \kepler{} space telescope on \hatpeleven{}, presented on Figure \ref{fig:recurrence}. Lightcurves of transits 217, 223, 229, and 235 exhibit very similarly shaped spot-induced anomalies. The striking similarity between these four anomalies, spaced apart by six planetary orbits, suggests that they are caused by the same spot, which evolves little during these observations. If this is indeed the case, then we are seeing the same \strobo{} as \citet{2011ApJS..197...14D} on \keplerseventeen{}, and the similarity of transit anomalies implies that the period ratio is very close to 6:1.

\begin{figure}
\begin{center}
\includegraphics*[width=89mm]{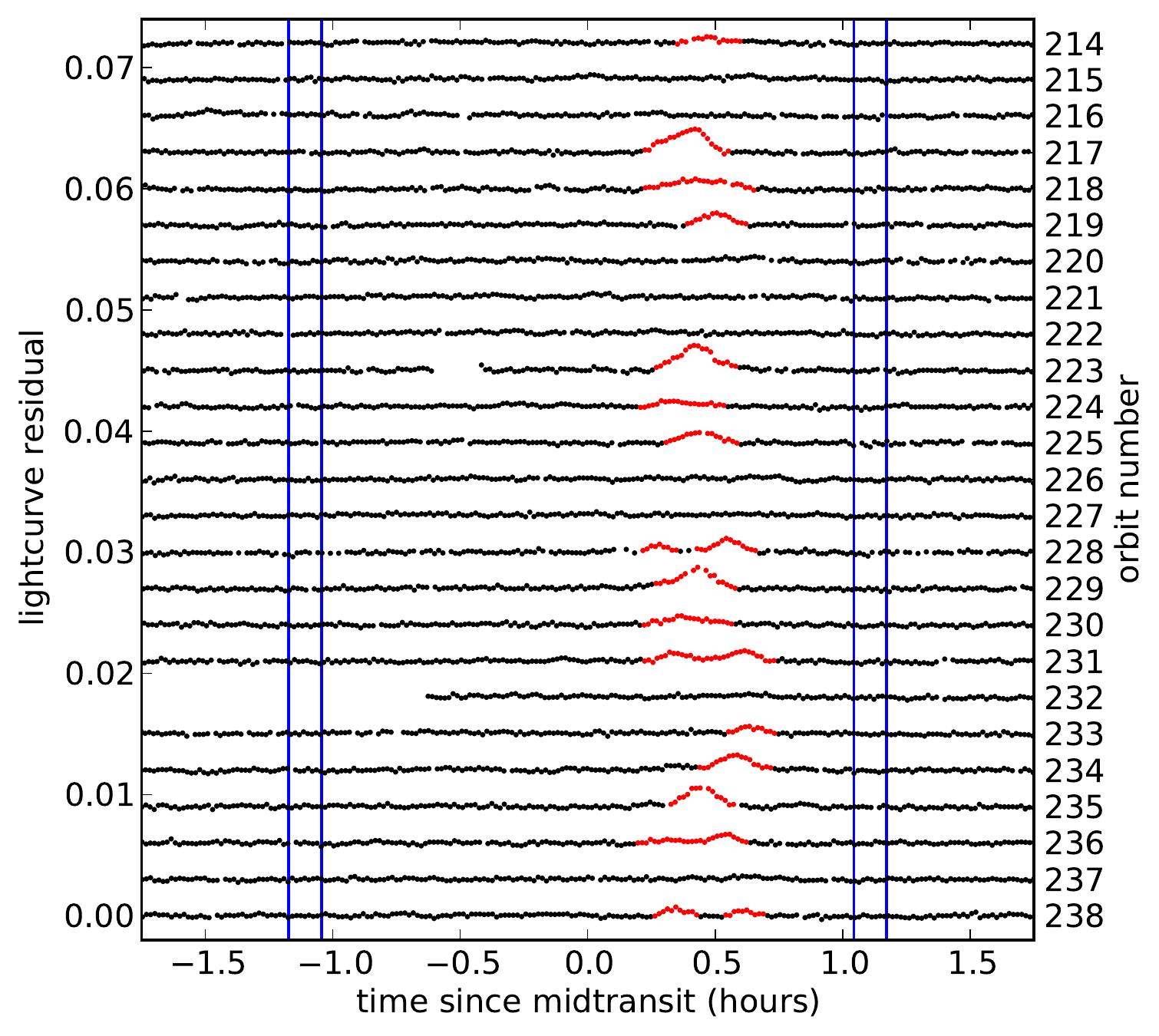}
\end {center}
\caption{\kepler{} short cadence observations minus Mandel--Agol model lightcurve for transits 214 to 238, as a function of time. Residuals are vertically displaced for each transit. Modelled spots (among \totalspots{} in total) are indicated with red points. The four blue vertical lines indicate first, second, third, and fourth contacts, from left to right.}
\label{fig:anomalies}
\end{figure}

However, the same transit anomaly might be caused by a continuous active band encircling the star along a constant latitude. In this case, the anomaly shape would not depend on how much the star rotates between each six transits, and thus would provide no information on a possible commensurability. To exclude this possibility, we look at all transits surrounding the ones highlighted on Figure \ref{fig:recurrence}. We subtract the model transit lightcurve without spots \citep{2002ApJ...580L.171M} from the observed data, and plot the residuals for each transit on Figure \ref{fig:anomalies}.

\change{We look for anomalies in adjacent transits that are similar to the one seen on Figure \ref{fig:recurrence} in transits 217, 223, 229, and 235.  However, these adjacent transits exhibit anomalies either with much smaller amplitude (in transits 218, 224, 228, 230, and 236), or at a different orbital phase (in transit 234), or none at all (in transits 216 and 222).  Therefore we can exclude the case of a continuous dark band around the star, because such a band would cause transit anomalies of comparable amplitude at the same phase in every single transit.}

\change{Note, however, that it is not possible to determine the exact shape of the spots based on transit anomalies that only scan the star along sparse transit chords, therefore the determination of stellar rotation period hinges on the assumed shape of the spots, circular in our case.  If, on the other hand, the spots were elongated in longitude, then the shape of the transit anomaly would not be sensitive to the stellar rotation rate, therefore the \strobo{} could be observed even for incommensurable periods.}

\change{Also note that there are signs of other spots evolving on Figure \ref{fig:anomalies}, for example, between transits 225 (one small spot), 231 (now split into two), and 237 (disappeared), that are also separated by one stellar rotation each.}

For our analysis, we adopt the revised transit ephemeris, planetary radius and orbital semi-major axis relative to the stellar radius, orbital inclination, and limb darkening parameters of \citet{2011ApJ...740...33D}. Their analysis accounts for eclipsed and uneclipsed spots, and relies on the orbital eccentricity and argument of periastron measurement of \citet{2010ApJ...710.1724B} using both RV data and Hipparcos parallax for \hatpeleven{}.  \change{To normalize each transit, we divide short cadence data by a linear fit to the out-of-transit observation within 0.12 days from the midtransit time.}

\section{Comparison to out-of-transit lightcurve}
\label{sec:macula}

When analyzing the lightcurve periodicity to find the stellar rotation rate, we assumed that the lightcurve is dominated by rotational modulation of spots (as opposed to, for example, stellar pulsation). As \citet{2010ApJ...710.1724B} noted, the rotational modulation amplitude is indeed consistent with expectations based on observations of other K dwarfs. In this section, we offer an independent method to confirm this hypothesis: we first identify a number of spots via transit anomalies, then we model the rotational modulation caused by these spots and compare it to the observed lightcurve.

\change{We adopt the analysis of \citet{2014spotrod}, who manually identify \totalspots{} spots in 130 transits in the \kepler{} dataset, and run MCMC analysis to explore the spot parameters.}  The model they use assumes the same quadratic limb darkening law for spots and the rest of the photosphere. It also assumes that spots are circular on the surface of the star, that is, elliptical in projection. 

The best fitting lightcurves for transits 217, 223, 229, and 235 are shown on Figure \ref{fig:recurrence}, involving a single, independent spot for each transit. The strikingly similar best fit position of the spot as shown on the right panels further supports the hypothesis of stellar rotation--planetary orbit commensurability. We also highlight data points that are considered to be part of a spot anomaly according to the best fit model in red on Figures \ref{fig:recurrence} and \ref{fig:anomalies}.

We feed the parameters of the spots derived from the transit anomalies into the rotational modulation model \macula{} \citep{2012MNRAS.427.2487K}. Since a long-lived stationary spot would be detected each stellar rotation (like the spot appearing in multiple transits above), we model each detected spot as if it lived for a single stellar rotation only, coming to life on the far side of the star half a rotation before we detect the transit anomaly it causes, and ceasing to exist half a stellar rotation later, also on the far side. Since we see \hatpeleven{} almost equator-on \change{\citep{2010ApJ...723L.223W,2011PASJ...63S.531H}}, every spot we model gets to the far side of the star half a stellar rotation after it is eclipsed by the planet. If the same spot causes another transit anomaly one stellar rotation later, we model it as a separate spot that is created when the first one dies. This is the simplest way of treating spot evolution: properties of a long-lived spot are described by piecewise constant functions, with the jumps happening when the spot is not in sight, resulting in a continuous model lightcurve. In this treatment, we do not have to investigate whether two transit anomalies separated by an integer number of stellar rotations are due to the same spot or different spots, since we treat them as separate spots in both cases.

For the \macula{} model, we adopt the projected obliquity and inclination distribution given by \citet{2011ApJ...743...61S} using as prior the results of \citet{2010ApJ...723L.223W} based on the \rmeffect{}.

\begin{figure}
\begin{center}
\includegraphics*[width=89mm]{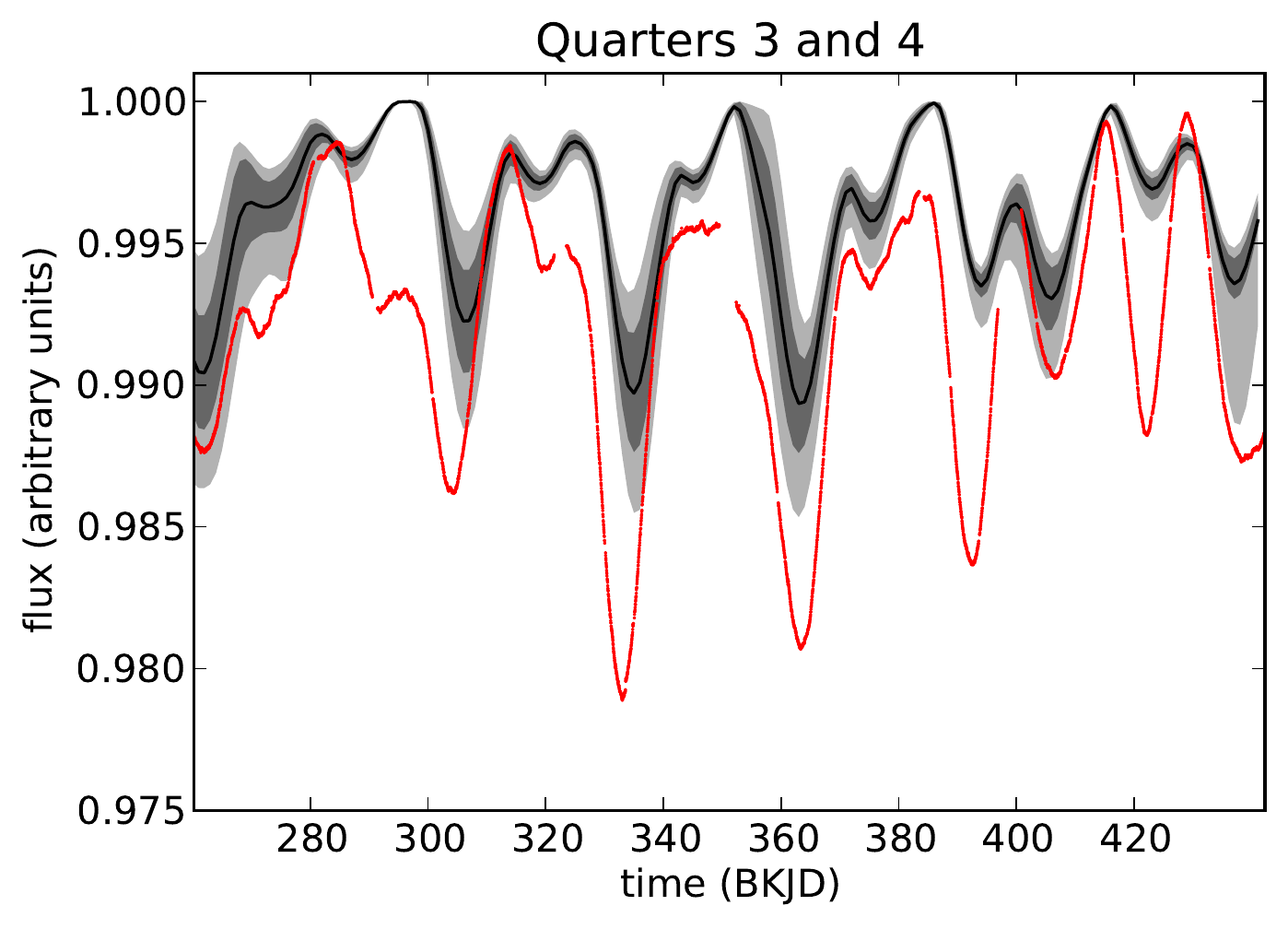}
\includegraphics*[width=89mm]{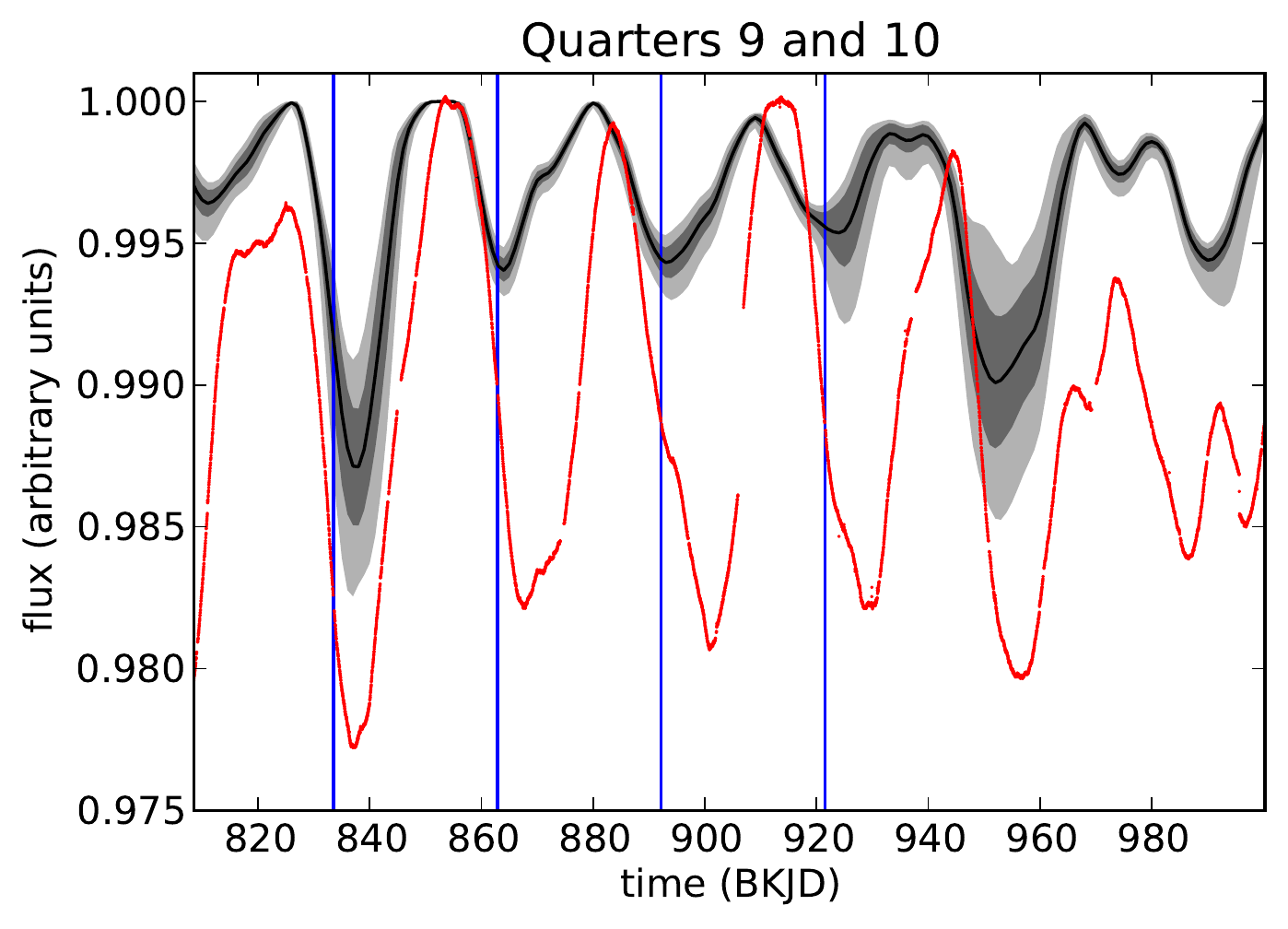}
\end {center}
\caption{Red dots: \kepler{} long cadence observations of \hatpeleven{} with hand-adjusted quarterwise scaling. Black curve and gray regions: \macula{} lightcurve model based on spots detected via transit anomalies, and its $1\sigma$ and $2\sigma$ confidence regions, accounting for the uncertainty in stellar inclination, projected obliquity, and spot parameters. This is not a fit for the out-of-transit lightcurve, but a model generated from spot parameters based on transit anomalies. Top panel shows quarters 3 and 4, bottom panels shows quarters 9 and 10. On the bottom panel, times of transits 217, 223, 229, and 235 are indicated by solid blue vertical lines.}
\label{fig:oot}
\end{figure}

Figure \ref{fig:oot} shows the long cadence observations in red, along with the \macula{} model lightcurve in black, for quarters 3, 4, 9, and 10. We also calculate the $1\sigma$ and $2\sigma$ confidence regions for the model lightcurve, accounting for the uncertainties of the inclination and projected obliquity as reported by \citet{2011ApJ...743...61S}, and the uncertainties of the spot parameters calculated from the transit anomalies.  \change{For the latter, we resample from the MCMC chains of \citet{2014spotrod}.}  The resulting confidence regions are highlighted in gray on the figure.

It is important to remember that this is not a fit for the out-of-transit lightcurve, but a model lightcurve using spot parameters inferred from a different phenomenon as input. We see that the model is a fair match to the observations in terms of qualitative features. \change{In particular, the deepest lightcurve minima are corretly predicted to occur after the transits drawn in blue on the bottom panel of Figure \ref{fig:oot}.  Projected obliquities of 90$^\circ$ and 270$^\circ$, both corresponding to a polar orbit, can be distinguised by the time of lightcurve minima, which would occur after or before the transit with the spot anomaly, respectively.  Our lightcurve analysis thus confirms the projected obliquity measurements based on the \rmeffect{}.}

However, the observed flux variations have an amplitude approximately 2 to 6 times larger than that of our model. It is likely that there are spots on the stellar surface that are never transited by the planet, therefore this model does not account for them. Such spots could contribute to the deeper minima in the observations, explaining the amplitude discrepancy.

For reference, the times of the four transits from Figure \ref{fig:recurrence} are also indicated on the bottom panel of Figure \ref{fig:oot} with blue vertical lines.

\change{Dark spots simultaneously increase transit depth and decrease the total brightness of the star \citep{2009A&A...505.1277C,2011ApJ...740...33D}.  Therefore we expect a negative correlation between these two quantities, albeit the variation in out-of-transit brightness and thus in transit depth is in the order of one percent, therefore the expected change of brightness during transit is a factor of few smaller than the noise of individual short cadence photometric data points.}

\change{To investigate this correlation, we calculate the out-of-transit brightness of \hatpeleven{} at the middle of each of the 204 transits by dividing the linear fit to short cadence out-of-transit data by mean intensity across the entire quarter.  We remove all data points that are affected by transit anomalies according to the best fit model, and fit a single transit depth scaling factor to the remaining data points, using a nominal Mandel--Agol lightcurve.}

\change{We find a Pearson correlation coefficient of $r=-0.20$ between out-of-transit intensity and transit depth, which does not indicate significant correlation.  In addition to the small expected variation of transit depth, we attribute this negative result to eclipsed spots that we do not identify during transits.  Note that an uneclipsed spot increases the transit depth, whereas an eclipsed one, when not accounted for, results in a shallower transit fit:  a bias in the opposite sense.  This potentially hinders the study of the correlation between out-of-transit brightness and transit depth.}

\section{Spot recurrence}
\label{sec:recurrence}

\citet{2014MNRAS.437.1045S} apply the method of hierarchical clustering to look for recurrence of spot anomalies in \keplerthirteenb{} transits, and they find a periodicity of three orbits with a high statistical significance. This implies that after three orbits, the planet rescans the same part of the stellar surface, supporting their hypothesis that an integer number of stellar rotations (five in this case) takes place during this time.

We aim to perform a statistical analysis of the same phenomenon on \hatpeleven{} in this section, using a different approach.  To analyze similarities between transits, we devise the following method: first, we calculate the deviation of the normalized transit lightcurves from the spotless model of \citet{2002ApJ...580L.171M}. Then we run a moving boxcar average of seven data points to decrease independent noise in the data. \change{After that, we set a threshold and flag transits with data points above it as anomalous. The next step is to pick a period and count pairs of observed transits that are spaced apart by this period. Finally, we plot the ratio of the ones among these pairs where both transits are flagged. If the planet could not eclipse the same spot in different transits, then anomalies would be independent, thus this ratio would not depend on the period.  In particular, if we flag $p$ fraction of total transits, then one randomly chosen transit is flagged with probability $p$, therefore two independent transits are simultaneously flagged with probability $p^2$ (as long as the number of transits is large).  Strong deviation of the ratio of flagged pairs of transits from $p^2$ as a function of period indicates correlated transit anomalies.}

Note that this method of identifying transit anomalies is different from manually picking them for fitting in Section \ref{sec:macula}. Using a uniform threshold has the advantage that detection does not rely on human decisions. We chose a large threshold (yielding fewer anomalies than what one can see by eye in the lightcurves) to avoid spurious detections.  It is important to note that the actual occurrence rates depend strongly on the choice of the threshold, although we find that the general features are persistent across a range of thresholds.

\begin{figure}
\begin{center}
\includegraphics*[width=89mm]{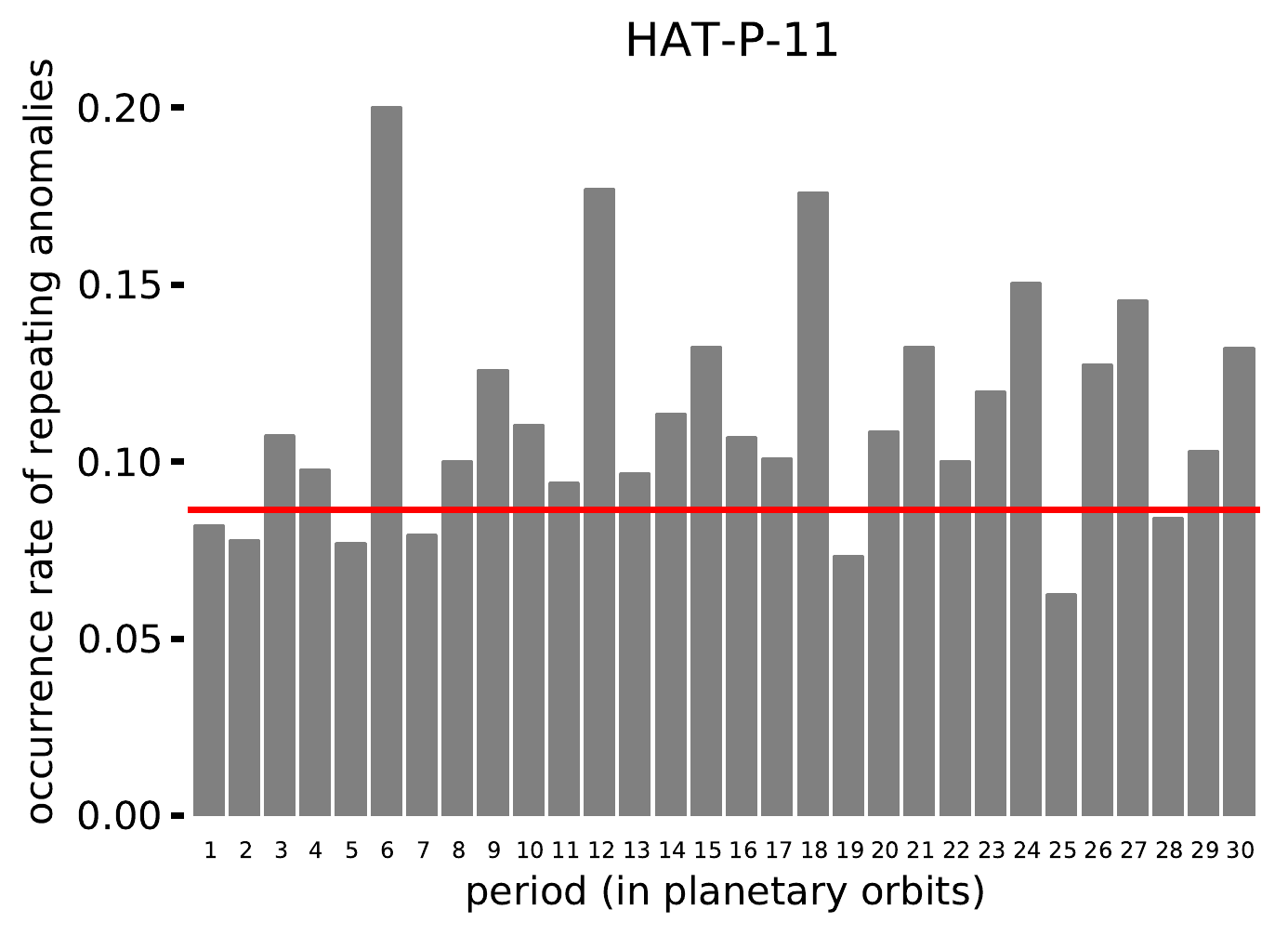}
\includegraphics*[width=89mm]{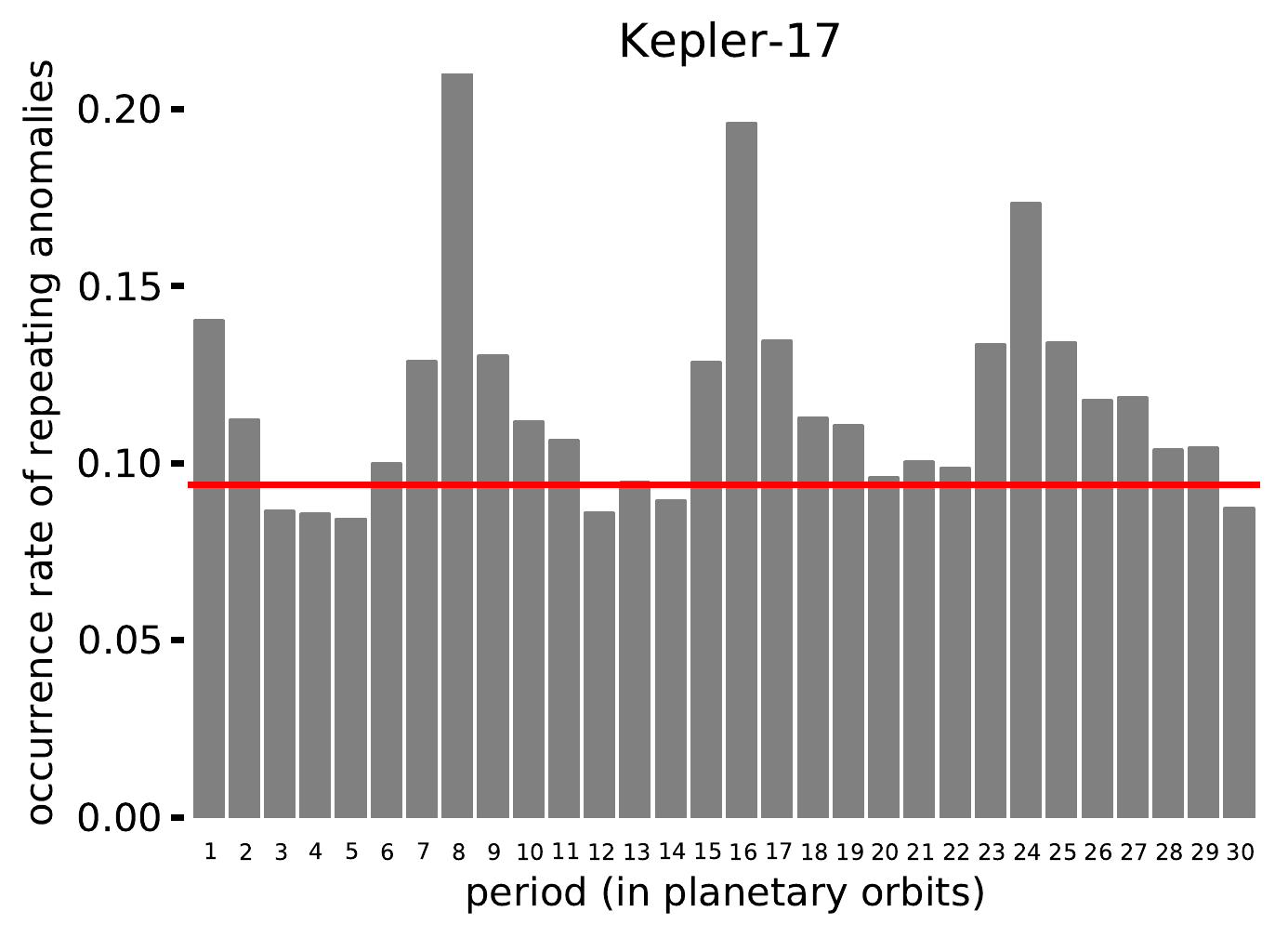}
\end{center}
\caption{Bars: occurrence fraction of transit pairs separated by a given number of orbits both exhibiting spot anomalies relative to the total number of pairs with the same spacing in the short cadence \kepler{} data, as a function of period. The ``statistical background'', the fraction expected if such spot anomalies were independent, is represented by red horizontal lines. Top panel is for \hatpeleven{}, bottom panel for \keplerseventeen{}.}
\label{fig:barchart}
\end{figure}

Good quality observations exist for 204 transits of \hatpelevenb{} in the \kepler{} dataset. We pick a threshold of $1\cdot10^{-4}$, which results in 60 flagged transits. That is, the occurrence rate of transit anomalies above this threshold is $p=0.29$.  \change{Figure \ref{fig:barchart} presents the ratios for a number of periods on the top panel, with the statistical background of $p^2=0.09$ overplotted as a horizontal red line.  For example, there are 165 pairs of observed transits that are six orbits apart.  If transits in each pair were flagged independently, we would expect to find $165p^2=14$ pairs of transits separated by six orbits with both transits flagged.  However, there are 33 such pairs in the dataset, more than two times as many.}

We perform the same analysis on 587 good quality transit lightcurves of \keplerseventeen{}. Since \keplerseventeen{} is a fainter target, we use a longer moving boxcar average, with 21 data points, to suppress photon noise. We use the same threshold as for \hatpeleven{}, resulting in 180 flagged transits. In this case, the occurrence rate is $p=0.31$, and the statistical background is $p^2=0.09$. The ratios of flagged pairs of transits as a function of period are presented on the bottom panel of Figure \ref{fig:barchart}.

We find the highest occurrence rate at periods of six and eight planetary orbits for \hatpeleven{} and \keplerseventeen{}, respectively: around 2.3 times the statistical background in both cases. We identify the next largest peaks as aliases of this frequency: at twelve, eighteen, and twenty-four orbits for \hatpeleven{}, and sixteen and twenty-four for \keplerseventeen{}. These are due to long-lived spots and exhibit decreasing strength, because not all spots that live for one stellar rotation continue to live for another one.

Based on these observations, we can exclude period aliases: if the star rotated two, three, or four times while the planet orbits six times, we would see a strong peak at three, two, or three orbital periods in case of \hatpeleven{} on Figure \ref{fig:barchart}, respectively. A similar argument holds for \keplerseventeen{}. These and higher harmonics can also be readily excluded based on the periodogram, the autocorrelation function, the projected rotational velocity of the star, and \apriori{} expectations of rotation rates \citep[see][for \hatpeleven{} and \keplerseventeen{}, respectively]{2010ApJ...710.1724B,2011ApJS..197...14D}.

An important difference between the interpretation of the results for the two planetary systems is due to their different geometry: \keplerseventeenb{} has an orbital axis well aligned with the projected spin axis of the star, therefore a spot would be eclipsed by the planet again even the periods were not commensurable. On the other hand, \hatpelevenb{} is known to be on a nearly polar orbit, therefore---as \citet{2010ApJ...723L.223W} pointed out---period commensurability is required for transit anomalies to recur, otherwise the planet would scan a different part of the stellar surface, missing the spot that it had eclipsed a stellar rotation earlier.

Another consequence of the orbital alignment of \keplerseventeenb{} with its host star's rotation is the excess on the side of each peak on the bottom panel of Figure \ref{fig:barchart}: at one, seven, nine, fifteen, seventeen, etc.~planetary orbits.  \change{As the star rotates, each spot seems to move parallel to the transit chord, thus spots are eclipsed in multiple subsequent transits \citep[see Figure 11 of][]{2011ApJS..197...14D}.  Therefore if a spot recurs eight, sixteen, twenty-four, etc.~planetary orbits later, it is likely to also cause an anomaly in the preceding and succeeding transits.}

On the other hand, transits of \hatpelevenb{} spaced apart by not an integer multiple of six orbits are expected to show spot-induced anomalies independently, because spots on the stellar surface rotate perpendicularly to the transit chord. This is indeed the case, except for secondary peaks at three, nine, fifteen, twenty-one and twenty-seven planetary orbits. The reason for these is the two opposite longitudes where spots seem to occur, as discussed in Section \ref{sec:flipflop}.

Finally, we note that the orbital period of \hatpelevenb{} is 3.3 times that of \keplerseventeenb{}, therefore periods in the upper panel represent correspondingly longer time than those in the lower panel. If we assume that spots have similar lifetime on the two stars, this explains why we see more noise for long periods for \hatpelevenb{} than for \keplerseventeenb{}.

\section{Flip-flop}
\label{sec:flipflop}

A lightcurve rotationally modulated by a single starspot has a well-defined minimum when the spot seems to be closest to the center of the stellar disk, and a flat maximum when the spot is behind the stellar limb. For a non-evolving spot, these minima happen repeatedly with the rotational period of the star. In this section, we make use of this effect, together with the assumption that lightcurve variations are mostly due to starspots (supported by the matching order of magnitude of amplitudes shown in Section \ref{sec:macula}), to confirm the rotational periods of \hatpeleven{} and \keplerseventeen{}.

\begin{figure}
\begin{center}
\includegraphics*[width=89mm]{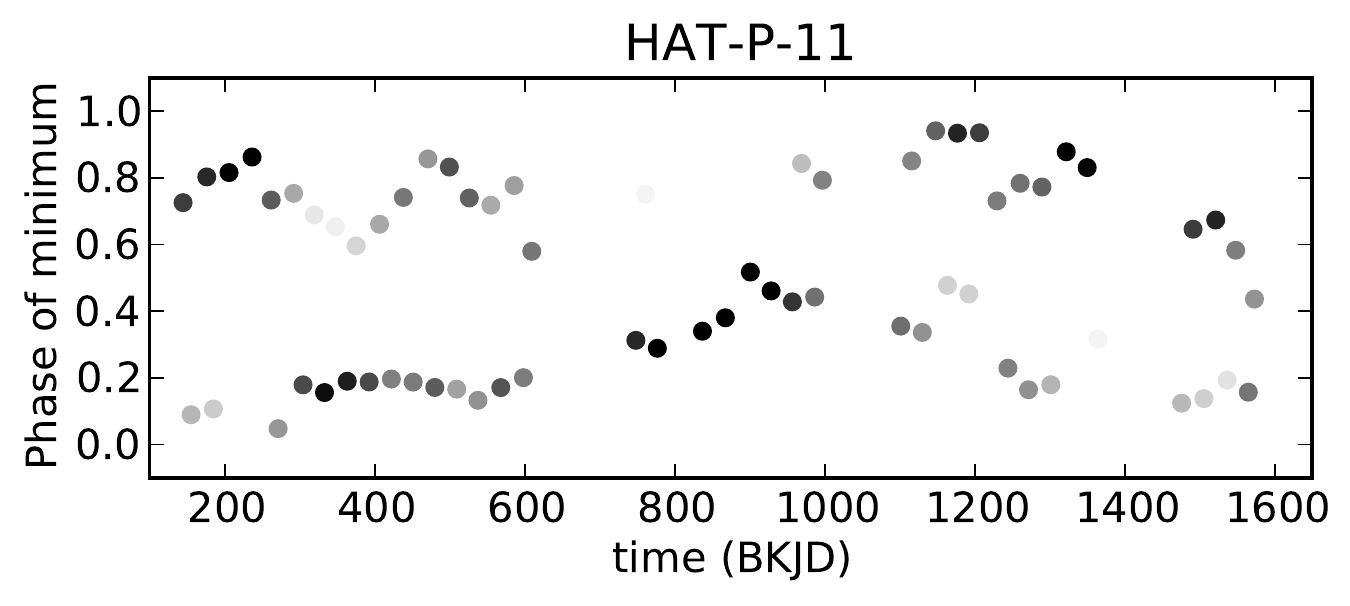}
\includegraphics*[width=89mm]{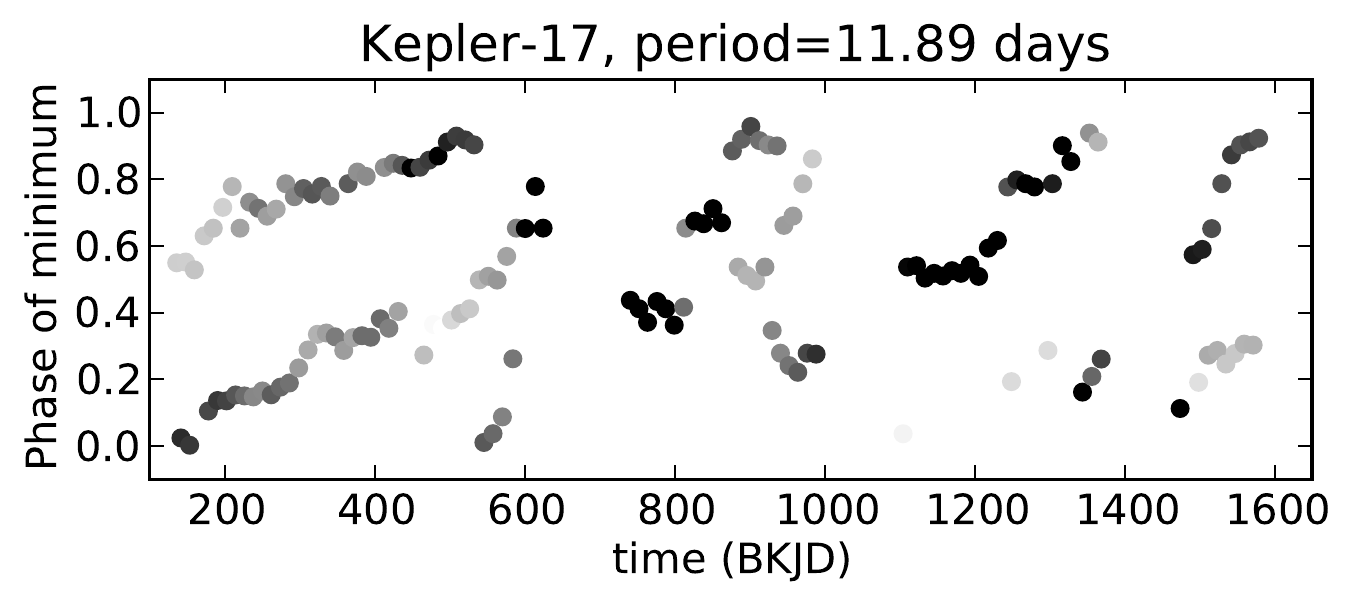}
\includegraphics*[width=89mm]{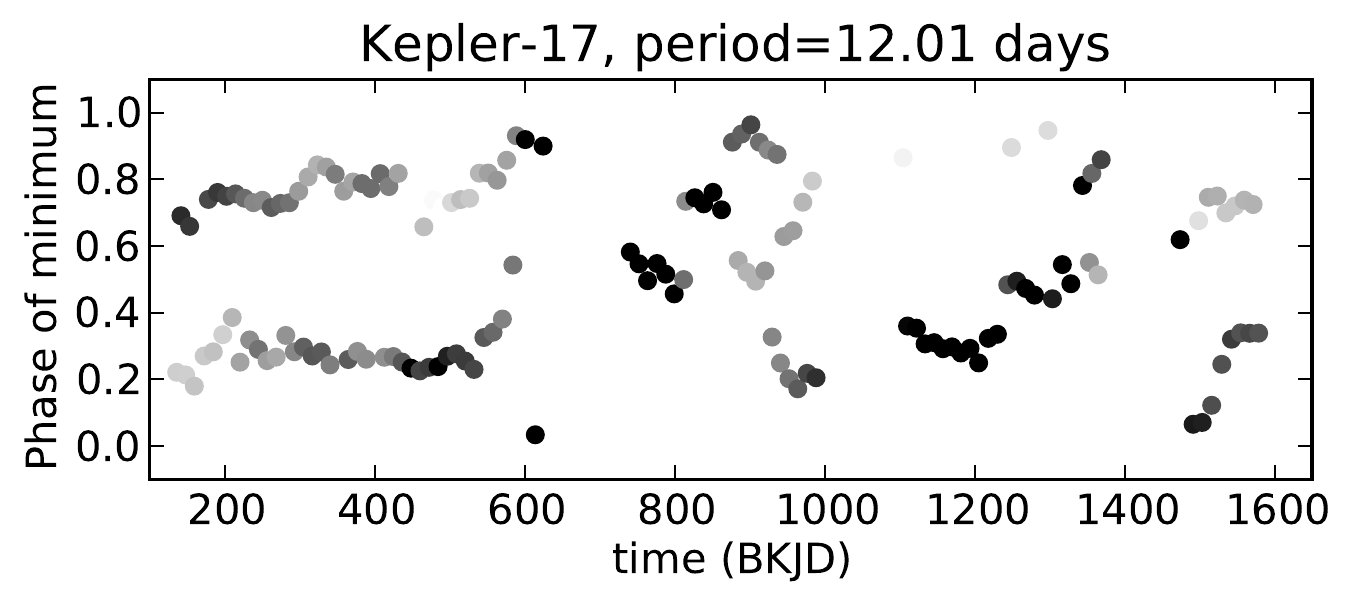}
\end{center}
\caption{Phases of lightcurve minima with respect to the proposed stellar rotation period as a function of time. The color of each filled circle corresponds to the difference between the minimum \change{brightness} and the \change{brightness} of the next smallest local minimum within 0.65 times the stellar rotation period: darker circles indicate relatively deeper minima. Top panel shows \hatpeleven{}, with the stellar rotation period being six times the planetary orbital period. The dominant phase changes from $0.7$ to $0.2$ around 300 BKJD, and it changes back around 1100 BKJD, which we interpret as flip-flop events. Middle panel shows \keplerseventeen{}, with minimum phases calculated using eight times the planetary orbital period as the stellar rotation period, as suggested by \citet{2011ApJS..197...14D}. Bottom panel shows \keplerseventeen{}, with stellar rotation period proposed by \citet{2012A&A...547A..37B}.}
\label{fig:flipflop}
\end{figure}

To this end, we identify local minima in their lightcurves. Figure \ref{fig:flipflop} shows the results for the two stars, indicating not only the time of each minimum on the horizontal axis, but also their phase relative to the stellar rotation on the vertical axis with the proposed stellar rotational period. For both stars, we find two minima during most stellar rotations, indicating two large spots (or spot groups) at opposite longitudes. This structure is responsible for the spurious signal in the periodograms at half the rotation period, on Figures \ref{fig:acf0} and \ref{fig:acf1}. On some other stars, this phenomenon might lead to incorrect identification of rotational periods \citep[see, for example,][]{2009MNRAS.400..451C}.

We find a very stable phase in case of \hatpeleven{} with the proposed stellar rotation period of six times the planetary orbital period, further supporting the proposed 6:1 commensurability (top panel). On the other hand, in case of \keplerseventeen{}, the phases of minima exhibit a large drift if we choose to calculate them with respect to eight times the planetary orbital period as the stellar rotation period (middle panel). This phase drift indicates that the real rotational period is different from what we used to calculate the phases. Indeed, we use the period 12.01 days as suggested by \citet{2012A&A...547A..37B} to recalculate the phases, and confirm that this yields phases of the minima without a significant drift (bottom panel). For both stars, the phase fluctuations of the minima might be due to spot migration, evolution, or new spots appearing at different longitudes.

The first discovery of phase jumps in a stellar lightcurve was reported by \citet{1991LNP...380..381J} on FK Comae Berenices, and named ``flip-flop behaviour''. \citet{2001A&A...379L..30K} attribute this phenomenon to two active regions on the star at opposite longitudes, with changing relative activity level. This results in minima in the lightcurve at two phases, with alternatingly one or the other being stronger. This phenomenon is exhibited by a large range of stars: RSCVn binaries, fast rotating G and K dwarfs, and the Sun \citep{2002AN....323..349H,2005LRSP....2....8B,2009A&ARv..17..251S}.

To determine whether a similar phenomenon takes place on the two stars we study, we quantify how much deeper each minimum is than the deepest neighboring minimum. On Figure \ref{fig:flipflop}, we represent minima that are much deeper than the ones half a stellar rotation earlier and later with black spots, and ones that are not so deep with lighter gray spots. Isolated minima, that is, ones that are not preceded or succeeded with one within less than one stellar rotation, are also black.

We interpret the results for \hatpeleven{} as evidence for two flip-flop events: the dominant phase changes from $0.7$ to $0.2$ around 300 BKJD, and it changes back around 1100 BKJD. The two year interval between these events is consistent with the flip-flop period on other stars \citep{2005LRSP....2....8B}. On the other hand, we are not able to interpret the results for \keplerseventeen{} as flip-flop cycles with a reasonable period.

\section{Star-planet interaction}
\label{sec:interaction}

A number of systems are known to exhibit commensurability between the planetary orbit and stellar rotation, for example \tauboo{} \change{\citep{2008A&A...482..691W}}, CoRoT-2 \change{\citep{2009EM&P..105..373P,2009A&A...493..193L}}, CoRoT-4 \change{\citep{2009A&A...506..255L}}, and \keplerthirteen{} \change{\citep{2012MNRAS.421L.122S}}.  This motivates the search for an underlying reason. To explore this phenomenon, first consider two systems without period commensurability. One is CoRoT-6, for which \citet{2011A&A...525A..14L} report that the spot covering factor gets enhanced when spots on the stellar surface cross a particular longitude with respect to the planet CoRoT-6b. \citet{2013A&A...553A..66H} find a similar behavior on the surface of LHS 6343 A, with a photospheric activity enhancement of existing spots, again at a particular position relative to its brown dwarf companion. 
Both research groups suggest that it is magnetic interactions that cause the enhancements; see, for example, the models of \citet{2006MNRAS.367L...1M,2006A&A...460..317P,2008A&A...487.1163L}.

A resonance effect might exists even if magnetic (or other) interactions between a planet and its host star were too weak to transfer enough angular momentum to make the planet migrate or to change the spin of the star. The same way the companions of CoRoT-6 and LHS 6343 A might cause an enhancement synchronous to their orbit, it is conceivable that if there was a latitude on the surface of a star with a period matching that of its companion, this effect, continuously acting on the same part of the stellar surface, would result in preferential spot formation at that latitude. For example, after measuring the differential rotation of \tauboo{}, \citet{2007MNRAS.374L..42C} note that this is such a system: the planetary orbital period falls between the stellar rotation periods at the equator and the pole, therefore there is an intermediate latitude with a period matching that of the planet. By this hypothesis, a relatively weak interaction might result in photospheric activity preferentially at this latitude, which would then cause photometric variations synchronous to the planetary orbit. Such variations were later detected by \citet{2008A&A...482..691W}, who indeed suggest magnetic interactions between the star and the planet as the cause of this phenomenon.

We extend this hypothesis from matching periods to general commensurability. For example, if the differential rotation profile of \hatpeleven{} happens to be such that at some intermediate latitude, the rotational period is exactly six times the orbital period of \hatpelevenb{}, then we propose that spot formation might be enhanced at this latitude by resonance with the planet, resulting in a lightcurve reflecting this commensurability, as we have shown in this paper.

It is possible that a number of small spots might form randomly at different latitudes on the surface of \hatpeleven{}. These spots might be too small to be detected through their contribution to rotational modulation or transit anomalies. We speculate that interactions with the planet might influence the growth or merger of such small spots, preferentially creating larger ones that we detect at resonant latitudes. Even though many of these large spots might only live for relatively few stellar rotations, the build-up phase during which large spots form from smaller ones might take much longer, possibly long enough for the hypothetical resonance to have a noticeable effect in the resulting large spot distribution as a function of latitude.

That we have presented another planetary system with a tight commensurability is not enough by itself to prove that there is an interaction at force between certain stars and their close-in planets. While \hatpeleven{} is a system with some very unique features, the tight 6:1 commensurability can still be purely by coincidence. One way to confirm our hypothesis is by a statistical analysis of a large number of planet hosts. One could compare the occurrence rate of detected commensurability to the rate predicted by our hypothesis using reasonable prior distributions for planetary orbital periods, stellar rotational periods, and differential rotation profiles. First of all, we note that this prediction is very sensitive to the assumed differential rotation profiles, and that differential rotation parameters have only been measured for relatively few stars. Another difficulty lies in the method of determining tight commensurability: as we have shown, neither a periodogram nor an autocorrelation function by itself is suitable for this. Part of the problem is the evolution of spots, and starspots occurring---although possibly in a smaller number---at other latitudes that do not have resonant rotational periods, thereby broadening the period peaks. The third method, using repeated transit anomalies like in this work or for \keplerseventeen{} by \citet{2011ApJS..197...14D}, is limited to bright host stars. Finally, identifying lightcurve minima and looking for their periodicity by itself might not be sufficient, since it is not clear \apriori{} if a star has active regions stationary on its surface, or the minima are due to spots appearing independently at random longitudes. One virtue of \hatpeleven{} is that we could apply and compare all these methods, and we are able to conclude that there are two dominant active regions, which seem to be stationary on the surface of the star for a long time.

We note that theoretically there are other ways to prove the hypothetical interaction between \hatpelevenb{} and its host star. \citet{2011ApJ...740...33D} point out that since the planet scans different latitudes of the stellar surface, it allows us to track the evolution of active latitudes with time. These observations would lead to a butterfly diagram, named after the characteristic migration pattern of active latitudes first observed on the Sun. \citet{2007ARep...51..675K} find similar behavior on some G and K dwarfs.

However, if interactions with \hatpelevenb{} induce preferential spot formation on \hatpeleven{} at fixed latitudes, this migration pattern might be suppressed. Therefore observing constant active latitudes instead of a butterfly-shaped migration pattern would be a strong indication for interactions between the planet and stellar surface activity. Unfortunately, the activity cycles for stars most similar to \hatpeleven{} in color and activity level as reported by \citet{1995ApJ...438..269B} span from 7 to 21 years (HD 201091, 190007, and 156026), which is much longer than the timespan of \kepler{} observations. Therefore even though we do not see strong evidence of spot migration in the \kepler{} data, we cannot yet determine whether the star exhibits Sun-like spot migration patterns on longer timescales.

It is also possible that active latitude migration is not suppressed on any star, no matter how strong the interaction with the planet is. For example, \citet{2007astro.ph..2530C} theoretically describe a mechanism that can cause the planetary interaction with the stellar magnetic fields to disappear at times (albeit for interactions with the chromosphere, not the photosphere). \citet{2008ApJ...676..628S} observationally confirm this phenomenon on both on HD 179949 and $\upsilon$ And, and dub it the \textit{on-off mechanism}. Even though there is no indication of such an event in the \kepler{} data for \hatpeleven{}, this possibility might make it impossible to confirm the effect of the planet based on spot migration patterns only.

\section{Conclusion}
\label{sec:conclusion}

The main focus of this paper is to present evidence for the 6:1 commensurability between the planetary orbit and the stellar rotation in the \hatpeleven{} system. For reference, we perform the same analysis for \keplerseventeenb{}, for which \citet{2011ApJS..197...14D} observe an 8:1 commensurability based on transit anomalies. However, \citet{2012A&A...547A..37B} show that in fact, spots with a different rotational rate dominate the out-of-transit lightcurve. These results are not necessarily contradictory because of possible differential rotation: in case of \keplerseventeen{}, the spots dominating the lightcurve might lie at a different latitude that the ones observed via anomalies in the transits of the planet with a low projected obliquity.

We calculate the autocorrelation function for the lightcurve of these two stars, and present a statistical analysis of possible spot-induced transit anomaly recurrence periods, which independently exclude frequency aliases of the proposed 6:1 and 8:1 commensurabilities. In case of \hatpeleven, the recurring transit anomalies imply a tight commensurability because of the polar orbit. We also present periodograms, and propose that the period discrepancy when looking at the FWHM of frequency peaks might be due to spot evolution causing the peaks to split.

We also present evidence for a tight 6:1 commensurability for \hatpeleven{} in the form of four observed transit anomalies presumably due to the same spot. We fit for all observed transit anomalies of \hatpeleven{}, and feed the resulting spot parameters into \macula{}, to show that it is plausible that rotational modulation accounts for most of the out-of-transit lightcurve variation. Furthermore, we identify minima in the lightcurve of both stars, and conclude that in case of \hatpeleven{}, there is a tight 6:1 period commensurability, whereas for \keplerseventeen{}, we confirm the period of 12.01 found by the much more sophisticated analysis of \citet{2012A&A...547A..37B}, distinct from the 8:1 commensurability. We identify two active longitudes for both stars, and see indication for two flip-flop events between these active longitudes on \hatpeleven{}.

Finally, we hypothesize that for stars with an intermediate latitude with a rotational period commensurable to the orbit of a close planet, star-planet interactions might induce spot formation preferentially at this latitude, which would show up as a resonance between the dominant period in the out-of-transit lightcurve and the planetary orbit, and also as the \strobo{} if the planet is transiting and the transit chord intersects this active latitude. However, proving this hypothesis might be difficult mostly because of the small number of bright targets and the uncertainties in differential rotation parameters.

\acknowledgements

Work by B.B.~and M.J.H.~was supported by NASA under grant NNX09AB28G from the Kepler Participating Scientist Program and grants NNX09AB33G and NNX13A124G under the Origins program. D.M.K.~is funded by the NASA Carl Sagan Fellowships. This paper includes data collected by the Kepler mission. Funding for the Kepler mission is provided by the NASA Science Mission directorate. The MCMC computations in this paper were run on the Odyssey 2.0 cluster supported by the FAS Science Division Research Computing Group at Harvard University. B.B.~is grateful for discussions with John A.~Johnson, Ruth Murray-Clay, and Claire Moutou. 

\bibliography{r}

\begin{thebibliography}{44}
\expandafter\ifx\csname natexlab\endcsname\relax\def\natexlab#1{#1}\fi

\bibitem[{{Aigrain} {et~al.}(2008){Aigrain}, {Collier Cameron}, {Ollivier},
  {Pont}, {Jorda}, {Almenara}, {Alonso}, {Barge}, {Bord{\'e}}, {Bouchy},
  {Deeg}, {de La Reza}, {Deleuil}, {Dvorak}, {Erikson}, {Fridlund}, {Gondoin},
  {Gillon}, {Guillot}, {Hatzes}, {Lammer}, {Lanza}, {L{\'e}ger}, {Llebaria},
  {Magain}, {Mazeh}, {Moutou}, {Paetzold}, {Pinte}, {Queloz}, {Rauer}, {Rouan},
  {Schneider}, {Wuchter}, \& {Zucker}}]{2008A&A...488L..43A}
{Aigrain}, S., {Collier Cameron}, A., {Ollivier}, M., {et~al.} 2008, \aap, 488,
  L43

\bibitem[{{Bakos} {et~al.}(2010){Bakos}, {Torres}, {P{\'a}l}, {Hartman},
  {Kov{\'a}cs}, {Noyes}, {Latham}, {Sasselov}, {Sip{\H o}cz}, {Esquerdo},
  {Fischer}, {Johnson}, {Marcy}, {Butler}, {Isaacson}, {Howard}, {Vogt},
  {Kov{\'a}cs}, {Fernandez}, {Mo{\'o}r}, {Stefanik}, {L{\'a}z{\'a}r}, {Papp},
  \& {S{\'a}ri}}]{2010ApJ...710.1724B}
{Bakos}, G.~{\'A}., {Torres}, G., {P{\'a}l}, A., {et~al.} 2010, \apj, 710, 1724

\bibitem[{{Baliunas} {et~al.}(1995){Baliunas}, {Donahue}, {Soon}, {Horne},
  {Frazer}, {Woodard-Eklund}, {Bradford}, {Rao}, {Wilson}, {Zhang}, {Bennett},
  {Briggs}, {Carroll}, {Duncan}, {Figueroa}, {Lanning}, {Misch}, {Mueller},
  {Noyes}, {Poppe}, {Porter}, {Robinson}, {Russell}, {Shelton}, {Soyumer},
  {Vaughan}, \& {Whitney}}]{1995ApJ...438..269B}
{Baliunas}, S.~L., {Donahue}, R.~A., {Soon}, W.~H., {et~al.} 1995, \apj, 438,
  269

\bibitem[{{Berdyugina}(2005)}]{2005LRSP....2....8B}
{Berdyugina}, S.~V. 2005, Living Reviews in Solar Physics, 2, 8

\bibitem[{{Bonomo} \& {Lanza}(2012)}]{2012A&A...547A..37B}
{Bonomo}, A.~S., \& {Lanza}, A.~F. 2012, \aap, 547, A37

\bibitem[{{Bord{\'e}} {et~al.}(2010){Bord{\'e}}, {Bouchy}, {Deleuil},
  {Cabrera}, {Jorda}, {Lovis}, {Csizmadia}, {Aigrain}, {Almenara}, {Alonso},
  {Auvergne}, {Baglin}, {Barge}, {Benz}, {Bonomo}, {Bruntt}, {Carone},
  {Carpano}, {Deeg}, {Dvorak}, {Erikson}, {Ferraz-Mello}, {Fridlund},
  {Gandolfi}, {Gazzano}, {Gillon}, {Guenther}, {Guillot}, {Guterman}, {Hatzes},
  {Havel}, {H{\'e}brard}, {Lammer}, {L{\'e}ger}, {Mayor}, {Mazeh}, {Moutou},
  {P{\"a}tzold}, {Pepe}, {Ollivier}, {Queloz}, {Rauer}, {Rouan}, {Samuel},
  {Santerne}, {Schneider}, {Tingley}, {Udry}, {Weingrill}, \&
  {Wuchterl}}]{2010A&A...520A..66B}
{Bord{\'e}}, P., {Bouchy}, F., {Deleuil}, M., {et~al.} 2010, \aap, 520, A66

\bibitem[{{Borucki} {et~al.}(2010){Borucki}, {Koch}, {Basri}, {Batalha},
  {Brown}, {Caldwell}, {Caldwell}, {Christensen-Dalsgaard}, {Cochran},
  {DeVore}, {Dunham}, {Dupree}, {Gautier}, {Geary}, {Gilliland}, {Gould},
  {Howell}, {Jenkins}, {Kondo}, {Latham}, {Marcy}, {Meibom}, {Kjeldsen},
  {Lissauer}, {Monet}, {Morrison}, {Sasselov}, {Tarter}, {Boss}, {Brownlee},
  {Owen}, {Buzasi}, {Charbonneau}, {Doyle}, {Fortney}, {Ford}, {Holman},
  {Seager}, {Steffen}, {Welsh}, {Rowe}, {Anderson}, {Buchhave}, {Ciardi},
  {Walkowicz}, {Sherry}, {Horch}, {Isaacson}, {Everett}, {Fischer}, {Torres},
  {Johnson}, {Endl}, {MacQueen}, {Bryson}, {Dotson}, {Haas}, {Kolodziejczak},
  {Van Cleve}, {Chandrasekaran}, {Twicken}, {Quintana}, {Clarke}, {Allen},
  {Li}, {Wu}, {Tenenbaum}, {Verner}, {Bruhweiler}, {Barnes}, \&
  {Prsa}}]{2010Sci...327..977B}
{Borucki}, W.~J., {Koch}, D., {Basri}, G., {et~al.} 2010, Science, 327, 977

\bibitem[{{Béky et al.}(2014)}]{2014spotrod}
{Béky et al.} 2014, {in preparation}

\bibitem[{{Catala} {et~al.}(2007){Catala}, {Donati}, {Shkolnik}, {Bohlender},
  \& {Alecian}}]{2007MNRAS.374L..42C}
{Catala}, C., {Donati}, J.-F., {Shkolnik}, E., {Bohlender}, D., \& {Alecian},
  E. 2007, \mnras, 374, L42

\bibitem[{{Collier Cameron} {et~al.}(2009){Collier Cameron}, {Davidson},
  {Hebb}, {Skinner}, {Anderson}, {Christian}, {Clarkson}, {Enoch}, {Irwin},
  {Joshi}, {Haswell}, {Hellier}, {Horne}, {Kane}, {Lister}, {Maxted}, {Norton},
  {Parley}, {Pollacco}, {Ryans}, {Scholz}, {Skillen}, {Smalley}, {Street},
  {West}, {Wilson}, \& {Wheatley}}]{2009MNRAS.400..451C}
{Collier Cameron}, A., {Davidson}, V.~A., {Hebb}, L., {et~al.} 2009, \mnras,
  400, 451

\bibitem[{{Cranmer} \& {Saar}(2007)}]{2007astro.ph..2530C}
{Cranmer}, S.~R., \& {Saar}, S.~H. 2007, ArXiv Astrophysics e-prints

\bibitem[{{Czesla} {et~al.}(2009){Czesla}, {Huber}, {Wolter}, {Schr{\"o}ter},
  \& {Schmitt}}]{2009A&A...505.1277C}
{Czesla}, S., {Huber}, K.~F., {Wolter}, U., {Schr{\"o}ter}, S., \& {Schmitt},
  J.~H.~M.~M. 2009, \aap, 505, 1277

\bibitem[{{Dawson} \& {Fabrycky}(2010)}]{2010ApJ...722..937D}
{Dawson}, R.~I., \& {Fabrycky}, D.~C. 2010, \apj, 722, 937

\bibitem[{{Deming} {et~al.}(2011){Deming}, {Sada}, {Jackson}, {Peterson},
  {Agol}, {Knutson}, {Jennings}, {Haase}, \& {Bays}}]{2011ApJ...740...33D}
{Deming}, D., {Sada}, P.~V., {Jackson}, B., {et~al.} 2011, \apj, 740, 33

\bibitem[{{D{\'e}sert} {et~al.}(2011){D{\'e}sert}, {Charbonneau}, {Demory},
  {Ballard}, {Carter}, {Fortney}, {Cochran}, {Endl}, {Quinn}, {Isaacson},
  {Fressin}, {Buchhave}, {Latham}, {Knutson}, {Bryson}, {Torres}, {Rowe},
  {Batalha}, {Borucki}, {Brown}, {Caldwell}, {Christiansen}, {Deming},
  {Fabrycky}, {Ford}, {Gilliland}, {Gillon}, {Haas}, {Jenkins}, {Kinemuchi},
  {Koch}, {Lissauer}, {Lucas}, {Mullally}, {MacQueen}, {Marcy}, {Sasselov},
  {Seager}, {Still}, {Tenenbaum}, {Uddin}, \& {Winn}}]{2011ApJS..197...14D}
{D{\'e}sert}, J.-M., {Charbonneau}, D., {Demory}, B.-O., {et~al.} 2011, \apjs,
  197, 14

\bibitem[{{Herrero} {et~al.}(2013){Herrero}, {Lanza}, {Ribas}, {Jordi}, \&
  {Morales}}]{2013A&A...553A..66H}
{Herrero}, E., {Lanza}, A.~F., {Ribas}, I., {Jordi}, C., \& {Morales}, J.~C.
  2013, \aap, 553, A66

\bibitem[{{Hirano} {et~al.}(2011){Hirano}, {Narita}, {Shporer}, {Sato}, {Aoki},
  \& {Tamura}}]{2011PASJ...63S.531H}
{Hirano}, T., {Narita}, N., {Shporer}, A., {et~al.} 2011, \pasj, 63, 531

\bibitem[{{Hussain}(2002)}]{2002AN....323..349H}
{Hussain}, G.~A.~J. 2002, Astronomische Nachrichten, 323, 349

\bibitem[{{Jetsu} {et~al.}(1991){Jetsu}, {Pelt}, {Tuominen}, \&
  {Nations}}]{1991LNP...380..381J}
{Jetsu}, L., {Pelt}, J., {Tuominen}, I., \& {Nations}, H. 1991, in Lecture
  Notes in Physics, Berlin Springer Verlag, Vol. 380, IAU Colloq. 130: The Sun
  and Cool Stars. Activity, Magnetism, Dynamos, ed. I.~{Tuominen}, D.~{Moss},
  \& G.~{R{\"u}diger}, 381

\bibitem[{{Katsova} {et~al.}(2007){Katsova}, {Bruevich}, \&
  {Livshits}}]{2007ARep...51..675K}
{Katsova}, M.~M., {Bruevich}, V.~V., \& {Livshits}, M.~A. 2007, Astronomy
  Reports, 51, 675

\bibitem[{{Kipping}(2012)}]{2012MNRAS.427.2487K}
{Kipping}, D.~M. 2012, \mnras, 427, 2487

\bibitem[{{Korhonen} {et~al.}(2001){Korhonen}, {Berdyugina}, {Strassmeier}, \&
  {Tuominen}}]{2001A&A...379L..30K}
{Korhonen}, H., {Berdyugina}, S.~V., {Strassmeier}, K.~G., \& {Tuominen}, I.
  2001, \aap, 379, L30

\bibitem[{{Lanza}(2008)}]{2008A&A...487.1163L}
{Lanza}, A.~F. 2008, \aap, 487, 1163

\bibitem[{{Lanza} {et~al.}(2009{\natexlab{a}}){Lanza}, {Pagano}, {Leto},
  {Messina}, {Aigrain}, {Alonso}, {Auvergne}, {Baglin}, {Barge}, {Bonomo},
  {Boumier}, {Collier Cameron}, {Comparato}, {Cutispoto}, {de Medeiros},
  {Foing}, {Kaiser}, {Moutou}, {Parihar}, {Silva-Valio}, \&
  {Weiss}}]{2009A&A...493..193L}
{Lanza}, A.~F., {Pagano}, I., {Leto}, G., {et~al.} 2009{\natexlab{a}}, \aap,
  493, 193

\bibitem[{{Lanza} {et~al.}(2009{\natexlab{b}}){Lanza}, {Aigrain}, {Messina},
  {Leto}, {Pagano}, {Auvergne}, {Baglin}, {Barge}, {Bonomo}, {Collier Cameron},
  {Cutispoto}, {Deleuil}, {de Medeiros}, {Foing}, \&
  {Moutou}}]{2009A&A...506..255L}
{Lanza}, A.~F., {Aigrain}, S., {Messina}, S., {et~al.} 2009{\natexlab{b}},
  \aap, 506, 255

\bibitem[{{Lanza} {et~al.}(2011){Lanza}, {Bonomo}, {Pagano}, {Leto}, {Messina},
  {Cutispoto}, {Moutou}, {Aigrain}, {Alonso}, {Barge}, {Deleuil}, {Fridlund},
  {Silva-Valio}, {Auvergne}, {Baglin}, \& {Collier
  Cameron}}]{2011A&A...525A..14L}
{Lanza}, A.~F., {Bonomo}, A.~S., {Pagano}, I., {et~al.} 2011, \aap, 525, A14

\bibitem[{{Mandel} \& {Agol}(2002)}]{2002ApJ...580L.171M}
{Mandel}, K., \& {Agol}, E. 2002, \apjl, 580, L171

\bibitem[{{McIvor} {et~al.}(2006){McIvor}, {Jardine}, \&
  {Holzwarth}}]{2006MNRAS.367L...1M}
{McIvor}, T., {Jardine}, M., \& {Holzwarth}, V. 2006, \mnras, 367, L1

\bibitem[{{McQuillan} {et~al.}(2013){McQuillan}, {Aigrain}, \&
  {Mazeh}}]{2013MNRAS.432.1203M}
{McQuillan}, A., {Aigrain}, S., \& {Mazeh}, T. 2013, \mnras, 432, 1203

\bibitem[{{Nutzman} {et~al.}(2011){Nutzman}, {Fabrycky}, \&
  {Fortney}}]{2011ApJ...740L..10N}
{Nutzman}, P.~A., {Fabrycky}, D.~C., \& {Fortney}, J.~J. 2011, \apjl, 740, L10

\bibitem[{{Pagano} {et~al.}(2009){Pagano}, {Lanza}, {Leto}, {Messina}, {Barge},
  \& {Baglin}}]{2009EM&P..105..373P}
{Pagano}, I., {Lanza}, A.~F., {Leto}, G., {et~al.} 2009, Earth Moon and
  Planets, 105, 373

\bibitem[{{Pont} {et~al.}(2007){Pont}, {Gilliland}, {Moutou}, {Charbonneau},
  {Bouchy}, {Brown}, {Mayor}, {Queloz}, {Santos}, \&
  {Udry}}]{2007A&A...476.1347P}
{Pont}, F., {Gilliland}, R.~L., {Moutou}, C., {et~al.} 2007, \aap, 476, 1347

\bibitem[{{Preusse} {et~al.}(2006){Preusse}, {Kopp}, {B{\"u}chner}, \&
  {Motschmann}}]{2006A&A...460..317P}
{Preusse}, S., {Kopp}, A., {B{\"u}chner}, J., \& {Motschmann}, U. 2006, \aap,
  460, 317

\bibitem[{{Rabus} {et~al.}(2009){Rabus}, {Alonso}, {Belmonte}, {Deeg},
  {Gilliland}, {Almenara}, {Brown}, {Charbonneau}, \&
  {Mandushev}}]{2009A&A...494..391R}
{Rabus}, M., {Alonso}, R., {Belmonte}, J.~A., {et~al.} 2009, \aap, 494, 391

\bibitem[{{Sanchis-Ojeda} \& {Winn}(2011)}]{2011ApJ...743...61S}
{Sanchis-Ojeda}, R., \& {Winn}, J.~N. 2011, \apj, 743, 61

\bibitem[{{Sanchis-Ojeda} {et~al.}(2011){Sanchis-Ojeda}, {Winn}, {Holman},
  {Carter}, {Osip}, \& {Fuentes}}]{2011ApJ...733..127S}
{Sanchis-Ojeda}, R., {Winn}, J.~N., {Holman}, M.~J., {et~al.} 2011, \apj, 733,
  127

\bibitem[{{Shkolnik} {et~al.}(2008){Shkolnik}, {Bohlender}, {Walker}, \&
  {Collier Cameron}}]{2008ApJ...676..628S}
{Shkolnik}, E., {Bohlender}, D.~A., {Walker}, G.~A.~H., \& {Collier Cameron},
  A. 2008, \apj, 676, 628

\bibitem[{{Silva}(2003)}]{2003ApJ...585L.147S}
{Silva}, A.~V.~R. 2003, \apjl, 585, L147

\bibitem[{{Silva-Valio}(2008)}]{2008ApJ...683L.179S}
{Silva-Valio}, A. 2008, \apjl, 683, L179

\bibitem[{{Strassmeier}(2009)}]{2009A&ARv..17..251S}
{Strassmeier}, K.~G. 2009, \aapr, 17, 251

\bibitem[{{Szab{\'o}} {et~al.}(2012){Szab{\'o}}, {P{\'a}l}, {Derekas}, {Simon},
  {Szalai}, \& {Kiss}}]{2012MNRAS.421L.122S}
{Szab{\'o}}, G.~M., {P{\'a}l}, A., {Derekas}, A., {et~al.} 2012, \mnras, 421,
  L122

\bibitem[{{Szab{\'o}} {et~al.}(2014){Szab{\'o}}, {Simon}, \&
  {Kiss}}]{2014MNRAS.437.1045S}
{Szab{\'o}}, G.~M., {Simon}, A., \& {Kiss}, L.~L. 2014, \mnras, 437, 1045

\bibitem[{{Walker} {et~al.}(2008){Walker}, {Croll}, {Matthews}, {Kuschnig},
  {Huber}, {Weiss}, {Shkolnik}, {Rucinski}, {Guenther}, {Moffat}, \&
  {Sasselov}}]{2008A&A...482..691W}
{Walker}, G.~A.~H., {Croll}, B., {Matthews}, J.~M., {et~al.} 2008, \aap, 482,
  691

\bibitem[{{Winn} {et~al.}(2010){Winn}, {Johnson}, {Howard}, {Marcy},
  {Isaacson}, {Shporer}, {Bakos}, {Hartman}, \&
  {Albrecht}}]{2010ApJ...723L.223W}
{Winn}, J.~N., {Johnson}, J.~A., {Howard}, A.~W., {et~al.} 2010, \apjl, 723,
  L223

\end{thebibliography}

\end{document}